\documentclass[iop]{emulateapj-rtx4}

\usepackage{mathrsfs }
\usepackage{natbib}
\usepackage{amssymb}
\usepackage{apjfonts}
\usepackage{ulem} 
\usepackage{graphicx}
\newcommand{\secpoint}{\mbox{$''\mskip-7.6mu.\,$}}

\begin{document}

\title{The First Robust Constraints on the Relationship Between Dust-to-Gas Ratio and Metallicity in Luminous Star-forming
Galaxies at High Redshift\altaffilmark{1}}

\author{
Alice E. Shapley,\altaffilmark{2}
Fergus Cullen,\altaffilmark{3}
James S. Dunlop,\altaffilmark{3}
Ross J. McLure,\altaffilmark{3}
Mariska Kriek,\altaffilmark{4}
Naveen A. Reddy,\altaffilmark{5}
Ryan L. Sanders\altaffilmark{6}
}

\altaffiltext{1}{Based on data obtained at the W.M. Keck Observatory, which is operated as a scientific partnership among the California Institute of Technology, the University of California,  and the National Aeronautics and Space Administration, and was made possible by the generous financial support of the W.M. Keck Foundation.}
\altaffiltext{2}{Department of Physics and Astronomy, University of California, Los Angeles, 430 Portola Plaza, Los Angeles, CA 90095, USA}
\altaffiltext{3}{Institute for Astronomy, University of Edinburgh, Royal Observatory, Edinburgh EH9 3HJ, UK}
\altaffiltext{4}{Astronomy Department, University of California at Berkeley, Berkeley, CA 94720, USA}
\altaffiltext{5}{Department of Physics and Astronomy, University of California, Riverside, 900 University Avenue, Riverside, CA 92521, USA}
\altaffiltext{6}{Department of Physics, University of California, Davis, 1 Shields Avenue, Davis, CA 95616, USA}
\email{aes@astro.ucla.edu}

\shortauthors{Shapley et al.}



\shorttitle{Dust-to-Gas-Ratio and Metallicity at High Redshift}

\begin{abstract}
We present rest-optical spectroscopic properties of a sample
of four galaxies in the Atacama Large Millimeter/submillimeter Array {\it Hubble}
Ultra Deep Field (ALMA HUDF). These galaxies span the redshift
range $1.41 \leq z \leq 2.54$ and the stellar mass
range $10.36\leq\log(M_*/{\rm M}_{\odot})\leq10.91$.
They have existing far-infrared and radio measurements
of dust-continuum and molecular gas emission from which bolometric star-formation
rates (SFRs), dust masses, and molecular gas masses have been estimated.
We use new $H$- and $K$-band near-infrared spectra from the
Keck/MOSFIRE spectrograph to estimate SFRs from dust-corrected H$\alpha$
emission (SFR(H$\alpha$)) and gas-phase oxygen abundances from the ratio 
[NII]$\lambda 6584$/H$\alpha$.
We find that the dust-corrected SFR(H$\alpha$)
is systematically lower than the bolometric SFR by a factor of several, and
measure gas-phase oxygen abundances in a narrow range,
$12+\log(\mbox{O/H})=8.59-8.69$ ($0.8-1.0\: (\mbox{O/H})_{\odot}$).
Relative to a large $z\sim 2$ comparison sample from the MOSDEF survey,
the ALMA HUDF galaxies scatter roughly symmetrically around the best-fit linear mass-metallicity relation,
providing tentative evidence for a flattening in the SFR dependence of metallicity
at high stellar mass. Combining oxygen abundances with
estimates of dust and molecular gas masses, we show
that there is no significant evolution in the normalization of the
dust-to-gas ratio DGR vs. metallicity relation from $z\sim0$ to $z\sim2$.
This result is consistent with some semi-analytic models and cosmological simulations describing
the evolution of dust in galaxies. Tracing the actual form of the DGR vs. metallicity relation
at high redshift now requires combined measurements of dust, gas, and
metallicity over a significantly wider range in metallicity.
\end{abstract}


\section{Introduction}
\label{sec:intro}
Dust is a key component of the interstellar medium (ISM), modulating interstellar
chemistry and thermodynamics. Dust also plays an important role
in the observed properties of galaxies over cosmic time, given that dust grains
absorb ultraviolet and optical starlight and reradiate it at longer wavelengths.
Indeed, our census of the star-formation rate (SFR) density of the universe
as a function of redshift is highly incomplete unless we include the fraction
of star formation obscured by dust \citep[e.g.,][]{madau2014}. Of particular
interest is the co-evolution of the dust and metal content of galaxies,
which constrains models of the formation and destruction of dust grains,
and the overall chemical enrichment of galaxies \citep[e.g.,][]{feldmann2015,popping2017,
mckinnon2017,li2019,hou2019}.

In particular, the dust-to-gas ratio (defined as dust mass divided by gas mass, hereafter
DGR) has been shown to scale with gas-phase oxgyen abundance
in the local universe \citep{remyruyer2014,devis2019}. Regardless of the specific
form used to fit this relationship (broken or single power-law), it is shown that
galaxies at lower metallicities ($12+\log(\mbox{O/H})<8.0-8.2)$
have a lower ratio of dust to metals than more metal-rich galaxies.
This scaling relationship can be interpreted in terms of dust grain
growth in the ISM as galaxies evolve from lower to higher metallicity \citep{popping2017}.
Both semi-analytic models and numerical simulations have been used to 
describe the form and evolution of the DGR vs. metallicity relationship
over a wide range of redshift. While \citet{popping2017}, \citet{li2019} and \citet{triani2020} predict
very little evolution in the DGR vs. metallicity relationship out to $z\sim6$,
\citet{hou2019} predict that the DGR should be detectably lower at fixed metallicity by
$z\sim2-3$. Although robust DGR and gas-phase metallicities have been
assembled for hundreds of $z\sim0$ galaxies, the corresponding measurements
at $z>1$ do not exist.

\begin{figure*}[ht!]
\centering
\includegraphics[width=0.7\linewidth]{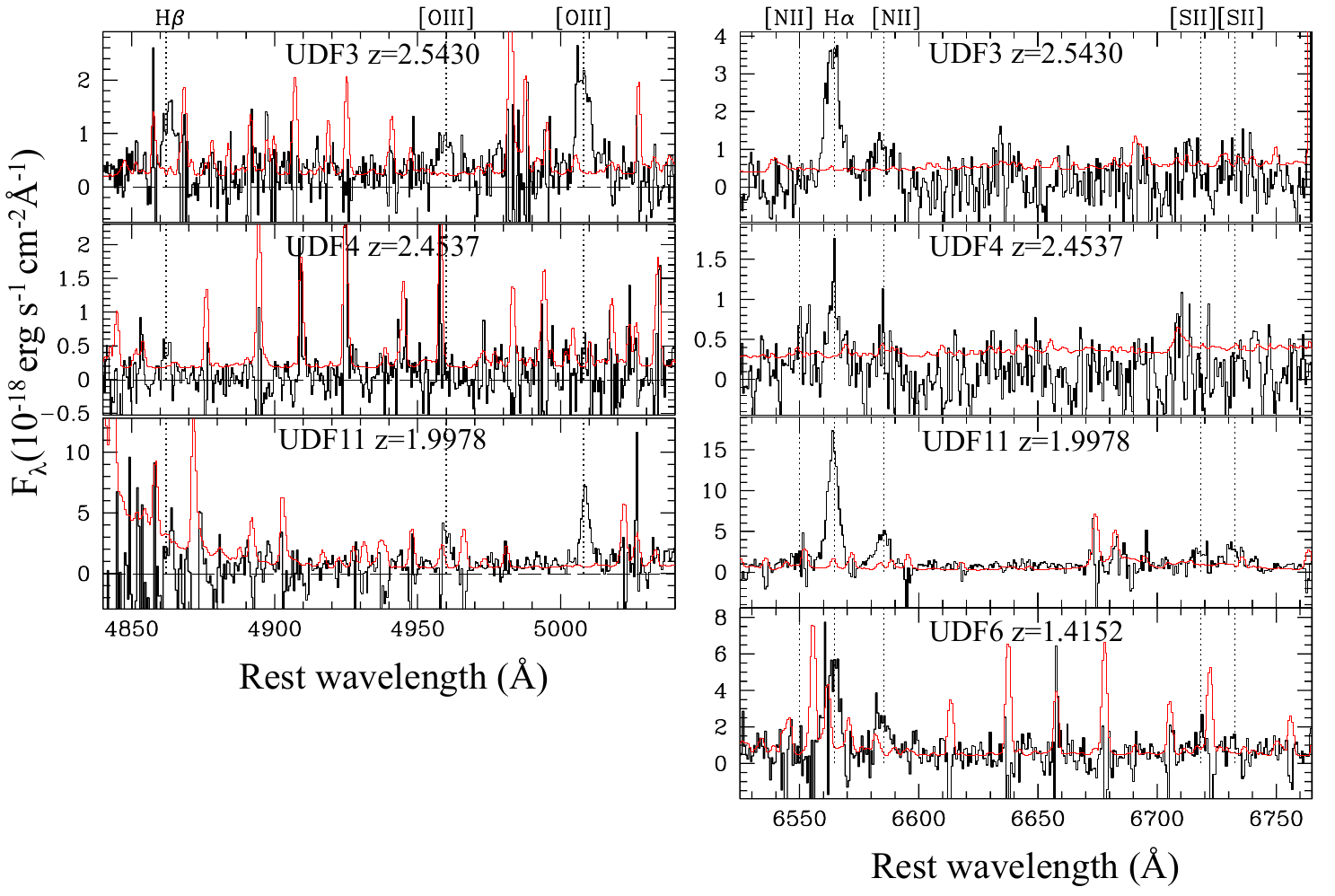}
\caption{Flux-calibrated MOSFIRE spectra for
the four targets in our sample. The spectra are shifted into the rest frame,
key nebular emission lines are labeled, and the $1\sigma$ error spectra are plotted in red.
The right-hand panels cover H$\alpha$, [NII]$\lambda6584$, and [SII]$\lambda\lambda6717,6731$
($K$-band for UDF3, UDF4, and UDF11 at $2.0\leq z\leq2.6$ and $H$-band for
UDF6 at $z=1.4152$), while the left-hand panels additionally cover H$\beta$ and [OIII]$\lambda5007$
in the $H$-band for UDF3, UDF4, and UDF11.}
\label{fig:mosfireplots}
\end{figure*}

Sensitive Atacama Large Millimeter/submillimeter Array (ALMA) and Karl G. Jansky Very Large Array (VLA)
observations of dust continuum and molecular
gas emission in the {\it Hubble} Ultra Deep Field \citep[HUDF;][]{dunlop2017,aravena2020,riechers2020}
enable DGR estimates for luminous star-forming galaxies out to $z\sim2-3$.
Here we present additional measurements of robust gas-phase metallicities based on rest-optical
spectroscopy, and therefore provide the first constraints on the DGR vs. metallicity
relationship in luminous star-forming galaxies at $z>1$.  We also compare dust-corrected rest-optical and bolometric measurements of SFR,
directly quantifying the importance of obscured star formation in our sample.
In Section~\ref{sec:obs}, we present the ALMA HUDF sample, and both existing and new observations.
Section~\ref{sec:results} contains our results on SFR, metallicity, and the DGR vs.
metallicity relation. We conclude in Section~\ref{sec:discussion}
with a discussion of the implications of these results.
Throughout, we assume a $\Lambda$CDM cosmology with
$H_0=70$ km s$^{-1}$ Mpc$^{-1}$, $\Omega_m=0.3$, and $\Omega_{\Lambda}=0.7$.

\section{Sample and Observations}
\label{sec:obs}

\subsection{ALMA HUDF Sample and Multi-wavelength Observations}
\label{sec:obs-almaudf}
We analyze a sample of four objects drawn from the ALMA HUDF 1.3~mm survey
of \citet{dunlop2017}, which used a 45-pointing mosaic to cover 
the 4.5 arcmin$^2$ region of the HUDF also imaged by the {\it Hubble} Wide Field Camera 3/IR.
A total of sixteen 1.3~mm detections were identified in the
ALMA HUDF, ranging in redshift from $z=0.67$ to $z=5.00$. This region was subsequently observed by the
ALMA Spectroscopic Survey in the {\it Hubble} Ultra Deep Field \citep[ASPECS;][]{walter2016}, which obtained
deeper continuum and higher-order CO emission-line observations at 1.2~mm \citep{aravena2020}, and additional
continuum and CO observations at 3.0~mm \citep{gonzalezlopez2019}. Furthermore, \citet{riechers2020}
performed VLA CO(1-0) observations at $\sim9$~mm in the ALMA HUDF.

The HUDF is covered by extensive multi-wavelength observations ranging from radio
to X-ray wavelengths. These include multi-wavelength photometry spanning from the visible
through mid-IR (i.e., {\it Spitzer}/IRAC) range, which can be used to model galaxy stellar populations. 
To estimate stellar masses, we model photometry from the publicly available catalogs of the 3D-HST 
survey \citep{skelton2014,momcheva2016}, corrected for contamination by rest-optical emission-line 
fluxes (Section~\ref{sec:obs-mosfire}). 

\subsection{MOSFIRE Observations}
\label{sec:obs-mosfire}
We used the Multi-object Spectrometer for Infrared Exploration
\citep[MOSFIRE;][]{mclean2012} on the Keck~I telescope to obtain moderate-resolution
$H$-, and $K$-band rest-optical spectra for the subset of targets from \citet{dunlop2017}
at spectroscopic or photometric redshifts where the strongest rest-optical nebular emission lines fall within windows of atmospheric
transmission ($1.4\leq z\leq1.7$, $2.0\leq z\leq2.6$, and $2.95\leq z\leq3.8$). There were
10 such targets, and we were able to fit 9 of them on a single multi-object slitmask 
centered at  R.A.=03:32:43.00 and decl.=$-$27:46:25.8 (J2000). We observed this mask
on 21 October 2018 and 13 January 2019, for a total of 3.2 hours ($64\times180$~seconds) in $K$ and 2.0 hours 
($59\times120$~seconds) in $H$. The slitwidth was 0\secpoint7, yielding a spectral resolution of $\sim 3650$
in $H$ and $\sim 3600$ in $K$. Conditions were partly cloudy on 21 October 2018 and clear
on 13 January 2019. The average seeing in both the $K$- and $H$-band spectra was 0\secpoint6.

\begin{figure}[ht!]
\centering
\includegraphics[width=\linewidth]{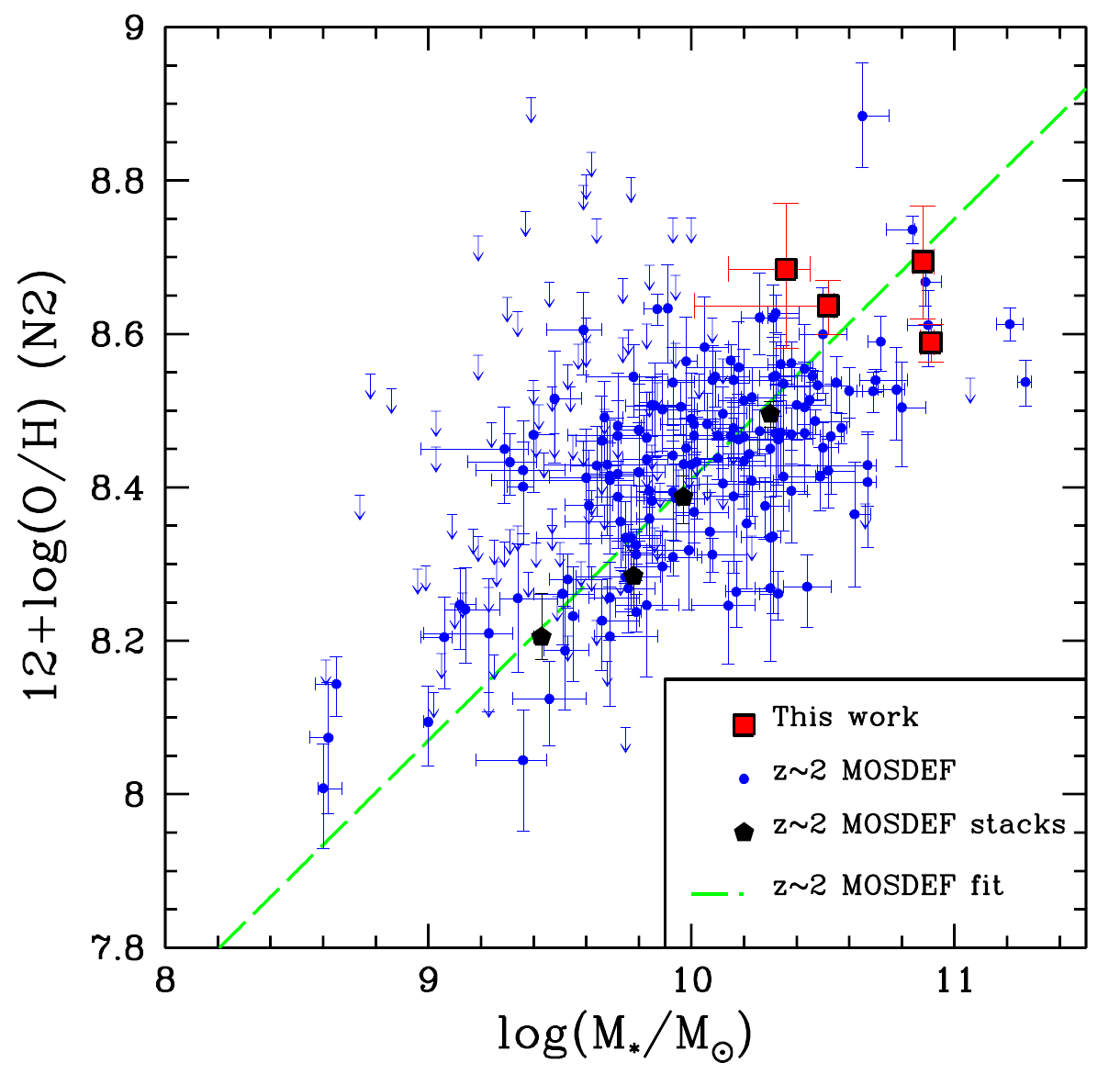}
\caption{Metallicity vs. stellar mass. Metallicity ($12+\log(\mbox{O/H})$)
is estimated from the N2 indicator. The ALMA HUDF targets are indicated
with red squares, while galaxies in the $z\sim2$ MOSDEF comparison sample
\citep{sanders2018} are indicated with blue circles (detections)
or downward-pointing arrows (upper limits). Stacks of $z\sim2$ MOSDEF galaxies
in bins of $M_*$ are shown with black pentagons. The fit to the MOSDEF stacks
is indicated with a green dashed line.}
\label{fig:mzr}
\end{figure}

We reduced the raw data to produce two-dimensional science and error spectra
using the pipeline described in \citet{kriek2015},
and optimally extracted one-dimensional science and error spectra from the two-dimensional spectra.
Flux calibrations and slit-loss corrections for each filter were applied as described in \citet{kriek2015}
and \citet{reddy2015}.
Of the 9 sources targeted with MOSFIRE, we measured rest-optical spectroscopic redshifts for five
(UDF3, UDF4, UDF6, UDF10, and UDF11), based on fitting Gaussian profiles to the strongest rest-optical emission lines. 
The remaining four sources, (UDF1, UDF7, UDF13, and UDF15) either have photometric redshifts at the edge of the
accessible ranges, such that their unmeasured spectroscopic redshifts may fall just outside the reach of MOSFIRE observations,
or CO redshifts \citep{gonzalezlopez2019,aravena2020} disagreeing with their photometric redshifts such that
rest-optical features fall outside windows of atmospheric transmission or in extremely noisy regions of the 
near-IR transmission windows.
The source UDF10 is not recovered in the deeper 1.2~mm map from
the ASPECS survey, and removed from further analysis, leaving a final sample of four sources.
Spectra for these sources are shown in Figure~\ref{fig:mosfireplots}, and their MOSFIRE
spectroscopic redshifts and H$\beta$, [OIII]$\lambda5007$, H$\alpha$, and [NII]$\lambda6584$
line fluxes are listed in Table~\ref{tab:mosfiremeas}.

\begin{deluxetable*}{lccccccrc}
\tabletypesize{\footnotesize}
\tablecolumns{9}
\tablecaption{MOSFIRE Measurements\label{tab:mosfiremeas}}
\tablehead{
\colhead{ID\tablenotemark{a}} & \colhead{3DHST-ID\tablenotemark{b}} & \colhead{R.A. (J2000)\tablenotemark{c}} & \colhead{Decl. (J2000)\tablenotemark{c}} & \colhead{$z_{\rm MOSFIRE}$\tablenotemark{d}} & \colhead{F(H$\beta$)} & \colhead{F([OIII]$\lambda5007$)} & \colhead{F(H$\alpha$)} & \colhead{F([NII]$\lambda6584$)}\\
\colhead{} & \colhead{(v4.1)} & \colhead{} & \colhead{} & \colhead{} & \colhead{($10^{-17}\frac{{\rm erg}}{{\rm s\: }{\rm cm}^2}$)} &  \colhead{($10^{-17}\frac{{\rm erg}}{{\rm s\: }{\rm cm}^2}$)} &  \colhead{($10^{-17}\frac{{\rm erg}}{{\rm s\: }{\rm cm}^2}$)} &  \colhead{($10^{-17}\frac{{\rm erg}}{{\rm s\: }{\rm cm}^2}$)}
}
\startdata
UDF3  & 29606 & 3:32:38.54 & $-$27:46:34.04 & 2.5430 & $1.61\pm0.34$ & $2.92\pm0.22$ & $7.38\pm0.40$  & $2.55\pm0.34$ \\
UDF4  & 29643 & 3:32:41.02 & $-$27:46:31.38 & 2.4537 & $0.43\pm0.13$ & $<1.71$       & $1.89\pm0.39$  & $0.79\pm0.24$ \\
UDF6  & 27881 & 3:32:34.43 & $-$27:46:59.55 & 1.4152 & \nodata       & \nodata       & $6.52\pm1.38$  & $2.84\pm0.59$ \\
UDF11 & 24110 & 3:32:40.05 & $-$27:47:55.44 & 1.9978 & \nodata       & $4.96\pm0.30$ & $18.60\pm0.47$ & $5.29\pm0.50$ 
\enddata
\tablenotetext{a}{Galaxy ID in \citet{dunlop2017}.}
\tablenotetext{b}{ID in the 3D-HST v4.1 catalog \citep{momcheva2016}.}
\tablenotetext{c}{Coordinates in the 3D-HST v4.1 catalog \citep{momcheva2016}.}
\tablenotetext{d}{Redshift from MOSFIRE spectra.}
\end{deluxetable*}

\section{Results}
\label{sec:results}
We combine our new MOSFIRE observations with the existing
multi-wavelength data in the ALMA HUDF to determine the relationships
among stellar populations, gas, dust and metals.

\subsection{Stellar Masses and SFRs}
\label{sec:res-sfr}

We estimated stellar masses by fitting photometry obtained from the
3D-HST photometric catalogs \citep{skelton2014,momcheva2016} with FAST++, 
an SED-fitting code written in C++ \footnote{https://github.com/cschreib/fastpp} 
and closely based on the FAST code \citep{kriek2009}. 
The \citet{conroy2009} flexible stellar
population synthesis models, a \citet{chabrier2003} initial mass
function (IMF), delayed-$\tau$ star-formation histories, and the \citet{calzetti2000} dust attenuation curve were
assumed. We estimated dust-corrected H$\alpha$ SFRs (SFR(H$\alpha$))
for UDF3 and UDF4, which have coverage of both H$\alpha$ and H$\beta$.
First we calculated $E(B-V)_{{\rm neb}}$, using the
stellar-absorption-corrected H$\alpha$/H$\beta$ Balmer decrement,
assuming  an intrinsic H$\alpha$/H$\beta$ emission-line ratio of 2.86
and the \citet{cardelli1989} extinction law. We then used $E(B-V)_{{\rm neb}}$
to dust-correct H$\alpha$ line fluxes. SFR(H$\alpha$) was 
then estimated from dust-corrected and slit-loss-corrected H$\alpha$ luminosities, based on the
calibration of \citet{hao2011} for a \citet{chabrier2003} IMF.
H$\beta$ falls in the $J$ band for UDF6, and in a region of low atmospheric
transmission for UDF11. Accordingly, we report uncorrected H$\alpha$ SFRs
as lower limits for these two galaxies.

An important question is how well dust-corrected H$\alpha$ emission traces
the bolometric SFR at high redshift. The answer likely depends on 
the bolometric SFR. While \citet{shivaei2016} has shown good agreement
between dust-corrected SFR(H$\alpha$) and bolometric SFR for a sample
of 17 star-forming galaxies at $z\sim2$ with SFR ranging from $\sim10-250\mbox{ M}_{\odot}{\rm yr}^{-1}$
(median 74$\mbox{ M}_{\odot}{\rm yr}^{-1}$), \citet{chen2020} show that
dust-corrected SFR(H$\alpha$) underpredicts the IR-based (i.e., bolometric) SFR
by at least a factor of 3 for a sample of 5 submillimeter galaxies
with a median IR-based SFR of $\sim500\mbox{ M}_{\odot}{\rm yr}^{-1}$.
We take ``total" SFRs from \citet{dunlop2017} (SFR$_{{\rm tot}}$) as the sum of UV (uncorrected
for dust) and far-IR SFRs, and compare them with dust-corrected SFR(H$\alpha$)
for UDF3 and UDF4 (shown in Table~\ref{tab:physical}). We find that dust-corrected SFR(H$\alpha$)
is a factor of 3.7$^{+2.7}_{-2.0}$ and 8.2$^{+13.0}_{-5.5}$ lower than SFR$_{{\rm tot}}$, respectively, for UDF3 and UDF4, although,
because of the large error bars on both sets of measurements, the difference is not highly
significant. If confirmed with smaller error bars, such a difference may indicate the contribution
of dusty star-forming regions that are opaque to H$\alpha$ emission but contribute to the bolometric
SFR in the far-IR.
UDF3 and UDF4, with SFR$_{{\rm tot}}$ values ranging from $\sim 100-200\mbox{ M}_{\odot}{\rm yr}^{-1}$,
appear consistent with the results of  \citet{chen2020}, although a larger sample of star-forming galaxies
spanning a wide range in SFR is needed, covered by measurements of 
H$\alpha$, H$\beta$ and IR-based tracers of bolometric luminosity.

\begin{deluxetable*}{lccccccccrcc}
\tabletypesize{\footnotesize}
\tablecolumns{12}
\tablewidth{7in}
\tablecaption{Derived Physical Properties\label{tab:physical}}
\tablehead{
\colhead{ID} & \colhead{$\log(M_*)$} & \colhead{SFR$_{\rm tot}$\tablenotemark{a}} & \colhead{SFR(H$\alpha$)\tablenotemark{b}} & \colhead{$12+\log({\rm O/H})$\tablenotemark{c}} & \colhead{$S_{1.2}$\tablenotemark{d}} & \colhead{$S_{3.0}$\tablenotemark{d}} & \colhead{$M_{\rm dust,1.2}$\tablenotemark{e}} & \colhead{$M_{\rm dust,3.0}$\tablenotemark{e}} & \colhead{$M_{\rm mol}$\tablenotemark{f}} & \colhead{DGR$_{\rm 1.2}$\tablenotemark{g}} & \colhead{DGR$_{\rm 3.0}$\tablenotemark{g}} \\[0.08cm]
\colhead{} & \colhead{(${\rm M}_{\odot}$)} & \colhead{($M_{\odot}\mbox{ yr}^{-1}$)} & \colhead{($M_{\odot}\mbox{ yr}^{-1}$)} & \colhead{(N2)} & \colhead{($\mu$Jy)} & \colhead{($\mu$Jy)} & \colhead{($10^8 {\rm M}_{\odot}$)} & \colhead{($10^8 {\rm M}_{\odot}$)} & \colhead{($10^{10} {\rm M}_{\odot}$)} & \colhead{($\times 10^{-3}$)} & \colhead{($\times 10^{-3}$)}
}
\startdata
UDF3 & 10.52$^{+0.02}_{-0.51}$ & 199.7$\pm$69.0 & 54.0$^{+42.0}_{-21.1}$ & 8.64$^{+0.03}_{-0.04}$ & 752$\pm$38 & 32$\pm$4 & 3.24$^{+0.64}_{-0.48}$ & 3.61$^{+0.62}_{-0.54}$ & 11.6$\pm$2.4 & 2.8$^{+1.0}_{-0.6}$ & 3.1$^{+1.0}_{-0.7}$ \\[0.10cm]
UDF4 & 10.36$^{+0.09}_{-0.22}$ & 94.4$\pm$4.0 & 11.5$^{+22.5}_{-7.0}$ & 8.68$^{+0.09}_{-0.10}$ & 316$\pm$16 & 23$\pm$4 & 1.39$^{+0.28}_{-0.20}$ & 2.61$^{+0.59}_{-0.52}$ & 2.4$\pm$0.9 & 5.8$^{+3.8}_{-1.7}$ & 10.9$^{+7.3}_{-3.5}$ \\[0.10cm] 
UDF6 & 10.88$^{+0.04}_{-0.04}$ & 87.1$\pm$11.0 & $>$3.7 & 8.69$^{+0.07}_{-0.07}$ & 430$\pm$23 & \nodata & 2.39$^{+0.40}_{-0.31}$ & \nodata & 10.0$\pm$0.8 & 2.4$^{+0.5}_{-0.3}$ & \nodata \\[0.10cm] 
UDF11 & 10.91$^{+0.01}_{-0.04}$ & 168.3$\pm$94.0 & $>$24.8 & 8.59$^{+0.02}_{-0.02}$ & 342$\pm$34 & $\leq 20$ & 1.69$^{+0.34}_{-0.26}$ & \nodata & 2.0$\pm$0.3 & 8.4$^{+2.3}_{-1.6}$ & \nodata 
\enddata
\tablenotetext{a}{``Total" SFR estimated from the sum of far-IR and UV emission, from Table~4 of \citet{dunlop2017}.}
\tablenotetext{b}{Dust-corrected H$\alpha$ SFR. For UDF3 and UDF4, the Balmer decrement is inferred from H$\alpha$ and H$\beta$. H$\beta$ is not covered for UDF6 and UDF11, so the uncorrected H$\alpha$ SFR is listed as a lower limit.}
\tablenotetext{c}{Gas-phase oxygen abundance based on the linear N2 indicator \citep{pettini2004}.}
\tablenotetext{d}{ALMA 1.2~mm and 3.0~mm fluxes from the ASPECS survey \citep{aravena2020}.}
\tablenotetext{e}{Dust mass in units of $10^8 M_{\odot}$, based on either $S_{1.2}$ or $S_{3.0}$, assuming an optically-thin modified blackbody function, and dust temperature, $T_{{\rm dust}}=35\pm5$K.}
\tablenotetext{f}{Molecular gas mass in units of $10^{10} M_{\odot}$. $M_{\rm mol}$ for UDF3 and UDF4 are taken from \citet{riechers2020}, based on CO(1-0) luminosities, while those for UDF6 and UDF11 come from \citet{aravena2020}, based on higher-order CO transitions.}
\tablenotetext{g}{Dust-to-gas ratio in units of $10^{-3}$, defined as the ratio between either $M_{\rm dust,1.2}$ or $M_{\rm dust,3.0}$ and $M_{\rm mol}$.}
\end{deluxetable*}

\subsection{Metallicities}
\label{sec:res-metals}
One of the key scaling relations probing the flow of baryons
through galaxies is the mass-metallicity
relation (MZR), between oxygen abundance and stellar mass.
The MZR has been traced from the local universe out
to $z\sim3$ \citep[e.g.,][]{tremonti2004,onodera2016}.
Since all four ALMA HUDF targets
have measurements of the [NII]$\lambda6584$/H$\alpha$ ratio, we 
use the N2 indicator \citep[N2$\equiv\log(\mbox{[NII]}\lambda6584/\mbox{H}\alpha$);][]{pettini2004} to estimate gas-phase oxygen abundance (Table~\ref{tab:physical}).
For the MZR analysis, we apply the linear form of the N2 calibration,
$12+\log(\mbox{O/H})=8.90+0.57\times\mbox{N2}$.
Our sample spans a narrow range of metallicity, from $\sim0.8-1.0$ solar,
\citep[assuming $12+\log(\mbox{O/H})_{\odot}=8.69$;][]{asplund2009}.
For context, in Figure~\ref{fig:mzr} we plot the masses and metallicities
of the ALMA HUDF sample alongside those of the $z\sim2$ star-forming
galaxy sample of \citet{sanders2018} from the MOSDEF survey.
$z\sim2$ MOSDEF galaxies ranging in mass from $\log({\rm M}_*/{\rm M}_{\odot})=9.0$
to $\log({\rm M}_*/{\rm M}_{\odot})=10.5$ are stacked in bins of ${\rm M}_*$. These
stacked points are plotted as well as their best-fit linear
regression \citep{sanders2018}.

The ALMA HUDF points lie at the massive end of the MOSDEF sample, with only UDF4
overlapping the most massive MOSDEF stack in ${\rm M}_*$, and
scatter around the best-fit linear MZR relation. Given the observed redshift evolution in the MZR,
we restrict the comparison to UDF3, UDF4, and UDF11, all at $z\sim 2$.
According to the ``Fundamental Metallicity Relation" \citep[FMR;][]{mannucci2010,andrews2013}
discovered among $z\sim0$ galaxies and 
describing the dependence of metallicity on both ${\rm M}_*$ and SFR, 
galaxies with {\it higher}-than-average SFR at fixed ${\rm M}_*$, 
lie at {\it lower}-than-average metallicity at fixed ${\rm M}_*$. \citet{sanders2018}
demonstrated that the FMR holds in this sense for $z\sim 2$ MOSDEF galaxies. 
In \citet{dunlop2017}, UDF3, UDF4, and UDF11 are shown to have bolometric SFR$_{{\rm tot}}$
that place them above the SFR vs. ${\rm M}_*$ main sequence on average, yet we observe 
no corresponding average offset below the MZR. 

At the same time, there is evidence that at 
the highest masses ($>10^{10.5}{\rm M}_{\odot}$)
in the local universe, the SFR dependence in the FMR reverses, such that galaxies
with higher-than-average SFRs have higher-than-average metallicities \citep{yates2012}. This high-mass regime,
in which most of the ALMA HUDF targets lie, is currently poorly sampled in the $z\sim 2$
MOSDEF dataset. A larger sample of high-mass $z\sim 2$ galaxies with metallicity
measurements spanning a wide range in bolometric SFRs is therefore needed to place the 
the masses, metallicities, and SFRs of the ALMA HUDF sample in context.
We mention one final possibility, which is AGN contamination
boosting the observed [NII]/H$\alpha$ ratios. UDF3 is identified as a radio-loud
AGN in \citet{dunlop2017}, yet its [OIII]$\lambda5007$/H$\beta$ and [NII]$\lambda6584$/H$\alpha$
ratios (Table~\ref{tab:mosfiremeas}) place it well within the distribution of $z\sim 2$ MOSDEF star-forming galaxies
in the [OIII]$\lambda5007$/H$\beta$ vs. [NII]$\lambda6584$/H$\alpha$ ``BPT" diagram \citep{shapley2019}.
Spatially-resolved integral-field unit spectroscopic observations may address
the question of AGN contamination.

\subsection{The Relationship Between DGR and Metallicity}
\label{sec:res-dgr}
The unique aspect of the dataset analyzed here is the combination of both dust and gas masses
drawn from the literature with new gas-phase oxygen abundance measurements. This combined
dataset enables us to investigate, for the first time, the relationship between DGR
and metallicity at $z>1$.

Given the greater depth of the ASPECS 1.2~mm observations, we use these measurements to infer dust masses 
instead of the original 1.3~mm measurements from \citet{dunlop2017}. For the two sources
in our sample with ASPECS 3.0~mm measurements (UDF3 and UDF4), we also use these longer-wavelength continuum
observations as independent proxies for dust mass. To translate 1.2 and 3.0~mm flux densities
($S_{1.2}$ and $S_{3.0}$) into dust masses, we assume an optically-thin modified
blackbody function \citep[e.g.,][]{hughes1997}, which yields:

\begin{equation}
M_{\rm dust} = \frac{S_{\nu}D_{\rm L}^2(z)}{\kappa_{\nu}B_{\nu}(T)(1+z)}
\label{eq:mdust}
\end{equation}

\bigskip
where $S_{\nu}$ is the observed-frame mm-wave flux density probing the Rayleigh-Jeans tail
of a modified blackbody; ${\kappa_{\nu}}$ is the dust mass absorption coefficient
at frequency, $\nu$, with a functional form ${\kappa_{\nu}}=\kappa_{850}(\frac{\nu}{\nu_{850\mu{\rm m}}})^{\beta}$,
and $\kappa_{850}$ is the opacity at 850~$\mu$m. Following recent work by \citet[e.g.,][]{liang2019},
we adopt $\kappa_{850}=0.05 \mbox{ m}^2\mbox{ kg}^{-1}$ and $\beta=2.0$. $B_{\nu}(T)$ is the Planck
function, and $D_{\rm L}(z)$ is the luminosity distance to redshift, $z$. For estimating dust mass,
we also assume a dust temperature of $T_{\rm dust}=35\pm5$K \citep{mclure2018}.

We further assume that the gas mass is well-approximated by the molecular gas mass, 
as is standard for $z\sim2$ star-forming galaxies \citep{tacconi2018},
and take $M_{{\rm mol}}$ from the literature. 
\citet{riechers2020} present $M_{{\rm mol}}$ for UDF3 and UDF4 based on CO(1-0) measurements,
while \citet{aravena2020} present $M_{{\rm mol}}$ for UDF6 and UDF11 based on higher-order CO transitions,
using the average CO excitation properties of the ASPECS sample \citep{boogaard2020} to convert to 
CO(1-0). In all cases, $\alpha_{{\rm CO}}=3.6 \mbox{ (K km s}^{-1}\mbox{pc}^2)^{-1}$ is assumed
to convert CO(1-0) luminosity to $M_{{\rm mol}}$. To estimate DGR, we simply take
the ratio of $M_{{\rm dust}}$ and $M_{{\rm mol}}$. $S_{1.2}$ and $S_{3.0}$, along with
$M_{{\rm dust}}$, $M_{{\rm mol}}$, and DGR, are listed in Table~\ref{tab:physical}.

We aim to compare the relationship between DGR and $12+\log(\mbox{O/H})$ for the ALMA HUDF
sample and the corresponding relationship in the local universe. Most recently,
\citet{devis2019} compiled metallicity and DGR measurements for 466 galaxies drawn
from the local Dustpedia sample \citep{davies2017}. $M_{{\rm dust}}$ for Dustpedia
galaxies was obtained from fitting multi-wavelength photometry spanning from the 
UV through the microwave. $M_{{\rm gas}}$ was estimated as the sum of 
atomic $M_{{\rm HI}}$ and molecular gas $M_{{\rm H2}}$, based on literature
measurements of $M_{{\rm HI}}$ and an assumed relationship
for $M_{{\rm H2}}/M_{{\rm HI}}$ as a function of $M_{{\rm HI}}/M_*$.
Using literature and VLT/MUSE optical emission-line spectra, \citet{devis2019}
determined global galaxy metallicities for a number of different strong-line metallicity
calibrations. These authors present best-fit linear regressions for DGR vs. metallicity
for each metallicity calibration, including N2 \citep{pettini2004}. In detail, \citet{devis2019}
used the cubic form of the N2 calibration, 
$12+ \log(\mbox{O/H})=9.37 +2.03\times\mbox{N2}+1.26\times\mbox{N2}^2+0.32\times\mbox{N2}^3$,
and fit a subsample of 368 late-type galaxies with $M_{{\rm dust}}$, $M_{{\rm gas}}$, and 
N2 measurements.
For a consistent comparison, we apply the cubic N2 calibration here to the ALMA HUDF sample,
which yields slightly higher $12+ \log(\mbox{O/H})$ values (on average 0.05~dex)
than the linear calibration within the N2 range spanned by the ALMA HUDF sample.
However, this comparison does not account for potential biases in the N2
metallicity calibration at high redshift due to evolving H~II region
physical conditions \citep{steidel2014,sanders2015,shapley2015}.

\begin{figure}[ht!]
\centering
\includegraphics[width=\linewidth]{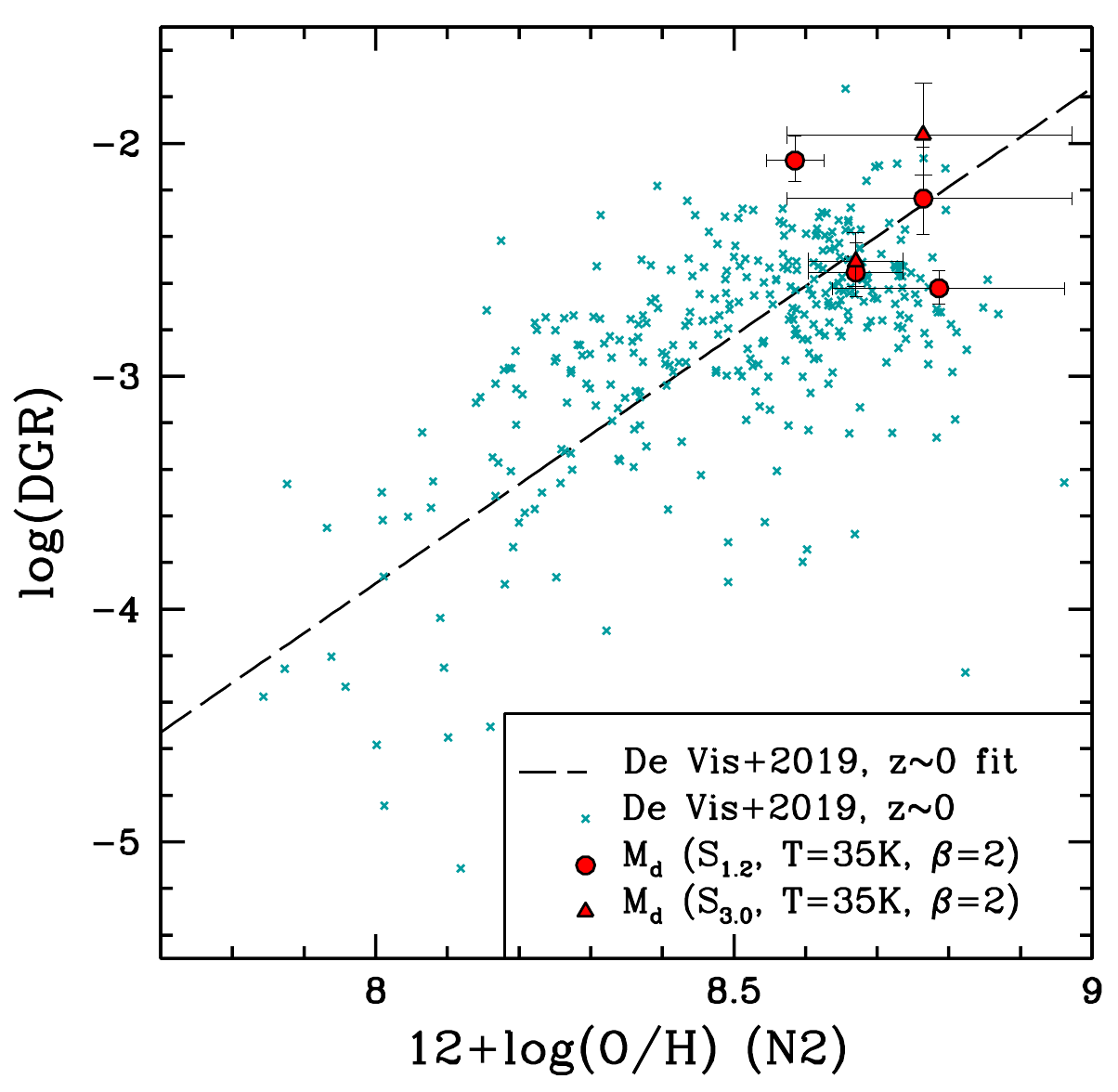}
\caption{Dust-to-gas ratio vs. $12+\log(\mbox{O/H})$. ALMA HUDF sources are indicated
with red symbols. Circles and triangles correspond to dust-to-gas ratios where $M_{{\rm dust}}$
was estimated, respectively, from the ALMA 1.2~mm continuum flux density
(all four sources) and 3.0~mm continuum flux density (UDF3 
and UDF4).  Turquoise crosses
indicate $z\sim0$ galaxies from \citet{devis2019}, while the dashed
line is the corresponding best-fit relation.  For both local and high-redshift
galaxies, plotted metallicities are estimated from the N2 indicator in its cubic form \citep{pettini2004}.}
\label{fig:dgr}
\end{figure}

As shown in Figure~\ref{fig:dgr}, we find that the DGRs for the
ALMA HUDF sample based on $S_{1.2}$ scatter around the local relation, with median DGR$_{\rm 1.2}$
within 0.04~dex of the DGR predicted by  \citet{devis2019} for the sample median metallicity.
DGR$_{\rm3.0}$ values for UDF3 and UDF4 are systematically higher, though consistent with the corresponding
DGR$_{\rm1.2}$ estimates. Unlike the local sample, the ALMA HUDF galaxies in the plot span only a very narrow range
in $12+\log(\mbox{O/H})$ (i.e., $\sim$0.2~dex, at roughly solar metallicity).
Accordingly, it is not possible to determine the slope
of the DGR vs. metallicity relation at $z\sim2$. However, 
at solar metallicity, there is robust evidence that the relationship between DGR and $12+\log(\mbox{O/H})$
remains constant from $z\sim0$ to $z\sim2$.

\section{Discussion}
\label{sec:discussion}

We have shown that the normalization of the DGR vs. metallicity relation at solar metallicity
does not significantly evolve between $z\sim0$ and $z\sim2$. Such a lack of evolution is consistent
with results at similar redshifts based on highly complementary measurements of the dust, gas, and metal content
of damped Ly$\alpha$ systems \citep{peroux2020} and star-forming regions probed
by gamma-ray-burst afterglow spectroscopy \citep{zafar2011}.
It is also consistent with the predictions of  the semi-analytic models of \citet{popping2017} and the
SIMBA cosmological simulations \citep{li2019}, in which galaxies evolve {\it along} the 
DGR vs. metallicity relationship as a function of time. At the same time, the simulations
of \citet{hou2019} predict measurable evolution towards lower DGR at fixed metallicity, as redshift
increases.  Accordingly, along with measures of the co-evolution of the dust and stellar mass
content of galaxies \citep[e.g.,][]{donevski2020},
evolutionary measures of the DGR vs. metallicity relation have discriminating
power among the descriptions of dust grain production and destruction over cosmic time in galaxy formation
simulations. Given the narrow range in metallicity
spanned by the ALMA HUDF sample,
however, we have no information on the form of the DGR vs. metallicity relation at $z\sim1-2$.
Future observations extending down towards lower metallicities and stellar masses -- and
correspondingly fainter ALMA dust continuum and CO flux levels -- will be required 
to determine the slope and scatter of the high-redshift DGR vs. metallicity relationship,
and compare with model predictions.

Assuming that not only the normalization but also the form of the DGR vs.
metallicity relationship remains constant to $z\sim2$, we can infer
the DGRs of high-redshift galaxies based on more easily obtained rest-optical spectroscopic measurements
of gas-phase metallicity. Such DGR estimates can be used to explain the relative
invariance out to $z\sim2$ in both the dust attenuation at 1600\AA\ ($A_{1600}$) and fraction
of star formation that is obscured ($f_{\rm obscured}$) at fixed 
galaxy ${\rm M}_*$ \citep[e.g.,][]{whitaker2017,mclure2018,cullen2018}.
The invariance of these relations is striking, given the strong redshift evolution
in galaxy properties at fixed ${\rm M}_*$ such as gas fraction and metallicity
\citep{sanders2018,tacconi2020}. These properties modulate dust attenuation,
which depends on both dust column density and wavelength-dependent opacity.
More quantitatively, we can express $A_{1600}$ as a function of
basic ISM properties as follows:

\begin{equation}
A_{1600} \propto \kappa_{1600} \times {\rm DGR} \times \Sigma_{{\rm gas}}
\label{eq:a1600}
\end{equation}

where $\kappa_{1600}$ is the dust opacity at 1600\AA\ in units of m$^2\mbox{ kg}^{-1}$,
and $\Sigma_{{\rm gas}}$ is the gas surface density, which can be inferred from inverting
the Kennicutt-Schmidt star-formation law. In future work, we will investigate whether lower
DGRs and higher $\Sigma_{{\rm gas}}$ at high redshift cancel out to keep  $A_{1600}$ and $f_{\rm obscured}$
roughly constant at fixed ${\rm M}_*$ -- or whether $\kappa_{1600}$, which encodes
the properties of dust grains themselves, must evolve as well to explain the observations.

\section*{Acknowledgements}
We acknowledge support from a NASA contract supporting the ``WFIRST Extragalactic Potential Observations (EXPO)
Science Investigation Team" (15-WFIRST15-0004), administered by GSFC,
and the support of the UK Science and Technologies Facilities Council.
We benefitted from useful conversations with Ian Smail, Desika Narayanan, and Qi Li.
We warmly acknowledge Pieter de Vis, for sharing the Dustpedia
dust-to-gas ratios and metallicity measurements.
We finally wish to extend special thanks to those of Hawaiian ancestry on
whose sacred mountain we are privileged to be guests.


\end{document}